\documentstyle[12pt,aaspp4,tighten,flushrt]{article}
\def\ti{\tilde}

\def\x{{\bf x}}
\def\part{\partial}

\def\bfd{{\bf d}}
\def\bbA{\overline{\bf A}}
\def\bA{\overline {A}}
\def\bfE{{\bf E}}
\def\bfJ{{\bf J}}
\def\bfH{{\bf H}}
\def\bfA{{\bf A}}
\def\bfB{{\bf B}}
\def\bbJ{\overline{\bf J}}

\def\bff{{\bf f}}

\def\BS{{\bf S}}

\def\bfa{{\bf a}}
\def\bfh{{\bf h}}

\def\lra#1{\left\langle #1\right\rangle}
\def\refp#1{(\ref{#1})}
\def\lab#1{\label{#1}}
\def\lrp#1{\left(#1\right)}
\def\lrb#1{\left[ #1\right]}
\def\cnt{\cdot\nabla\times}
\def\beqn{\begin{eqnarray}}
\def\eeqn{\end{eqnarray}}
\def\nn{\nonumber\\ }
\def\ni{\noindent}
\def\.{\mathaccent 95}
\def\beq{\begin{equation}}
\def\ee{\end{equation}}
\def\a{{\alpha}}
\def\be{{\beta}}
\def\ga{\gamma}
\def\de{\delta}
\def\ep{\epsilon}

\def\la{\lambda}

\def\si{\sigma}

\def\Om{\Omega}

\def\prt{\partial}

\def\frac#1#2{{\textstyle{{#1}\over {#2}}}}
\def\ni{\noindent}
\def\lsim{\mathrel{\rlap{\lower4pt\hbox{\hskip1pt$\sim$}}
    \raise1pt\hbox{$<$}}}
\def\gsim{\mathrel{\rlap{\lower4pt\hbox{\hskip1pt$\sim$}}
    \raise1pt\hbox{$>$}}}
\def\sqr#1#2{{\vcenter{\vbox{\hrule height.#2pt
         \hbox{\vrule width.#2pt height#1pt \kern#1pt
         \vrule width.#2pt}
         \hrule height.#2pt}}}}

\newbox\grsign \setbox\grsign=\hbox{$>$} \newdimen\grdimen \grdimen=\ht\grsign
\newbox\simlessbox \newbox\simgreatbox
\setbox\simgreatbox=\hbox{\raise.5ex\hbox{$>$}\llap
     {\lower.5ex\hbox{$\sim$}}}\ht1=\grdimen\dp1=0pt
\setbox\simlessbox=\hbox{\raise.5ex\hbox{$<$}\llap
     {\lower.5ex\hbox{$\sim$}}}\ht2=\grdimen\dp2=0pt

\def\doublespace {\smallskipamount=6pt plus2pt minus2pt
                  \medskipamount=12pt plus4pt minus4pt
                  \bigskipamount=24pt plus8pt minus8pt
                  \normalbaselineskip=24pt plus0pt minus0pt
                  \normallineskip=2pt
                  \normallineskiplimit=0pt
                  \jot=6pt
                  {\def\smallskip {\vskip\smallskipamount}}
                  {\def\medskip   {\vskip\medskipamount}}
                  {\def\bigskip   {\vskip\bigskipamount}}
                  {\setbox\strutbox=\hbox{\vrule 
                    height17.0pt depth7.0pt width 0pt}}
                  \parskip 12.0pt
                  \normalbaselines}

\font\gkvec=cmmib10                         
\def\bomega{\hbox{{\gkvec\char33}}}                  

\def\lb{\langle}
\def\rb{\rangle}
\def\bbE{\overline {\bf E}}
\def\bw{\overline{\omega}}
\def\bv{\overline V}
\def\bB{\overline B}
\def\ts{\times}
\def\lb{\langle}
\def\rb{\rangle}
\def\curl{\nabla {\ts}}

\def\cd{\cdot}
\def\bbV{\overline {\bf V}}
\def\bfv{{\bf v}}

\def\bfV{{\bf V}}
\def\bfj{{\bf j}}

\def\bfe{{\bf e}}
\def\bfw{{\bomega}}

\def\bfb{{\bf b}}

\def\bfB{{\bf B}}

\def\bbB{\overline{\bf B}}
\def\nb{\nabla}
\def\curl{\nb\ts}

\def\b0{b^{(0)}}
\def\v0{v^{(0)}}
\def\w0{\omega^{(0)}}
\def\bb0{\bfb^{(0)}}
\def\bv0{\bfv^{(0)}}
\def\bw0{\bfw^{(0)}}
\def\bj0{\bfj^{(0)}}

\def\ni{\noindent}
\begin{document}

\centerline{\bf Constraints on the magnitude of $\a$ in dynamo theory}
\medskip
\centerline{Eric G. Blackman} 
\centerline {Theoretical Astrophysics, Caltech 130-33, Pasadena CA, 91125, USA}
\centerline {and}
\centerline {George B. Field} 
\centerline {Center for Astrophysics, 60 Garden St., Cambridge MA, 02139, USA}
\centerline{(accepted to ApJ)}

\centerline {\bf ABSTRACT}
 
We consider the backreaction of the magnetic field
on the magnetic dynamo coefficients  
and the role of boundary conditions in interpreting
whether numerical evidence for suppression is dynamical.
If a uniform field in a periodic box 
serves as the initial condition for 
modeling the backreaction on the turbulent EMF, then
the magnitude of the turbulent EMF and thus the
dynamo coefficient $\a$, have a stringent upper limit 
that depends on the magnetic Reynolds number $R_M$ to a power of order $-1$.  
This is not a dynamic suppression but results 
just because of the imposed boundary conditions.
In contrast, when mean field gradients are allowed within the 
simulation region, or non-periodic boundary are used, 
the upper limit is independent of $R_M$ and takes its kinematic value.
Thus only for simulations of the latter types 
could a measured suppression be the result of a dynamic backreaction.
This is fundamental for understanding a long-standing controversy 
surrounding $\alpha$ suppression.  Numerical simulations which do not allow 
any field gradients and invoke periodic boundary conditions 
appear to show a strong $\alpha$ suppression (e.g. Cattaneo \& Hughes 1996).
Simulations of accretion discs which allow field gradients 
and allow free boundary conditions (Brandenburg \& Donner 1997) suggest
a dynamo $\alpha$ which is not suppressed by a power of $R_M$.
Our results  are consistent with {\it both} types of simulations.


\medskip

{\bf Subject Headings}: magnetic fields; galaxies: magnetic fields;
Sun: magnetic fields; stars: magnetic fields; turbulence; accretion discs.

\vfill
\eject
\centerline {\bf 1. Introduction}

A leading candidate to explain
the origin of large scale magnetic fields in stars and galaxies 
is mean-field turbulent magnetic dynamo theory 
(Moffatt 1978; Parker
1979; Krause \& R\"{a}dler 1980; Zeldovich et al.\ 1983, Ruzmaikin et al.\
1988, Beck et al.\ 1996). The theory appeals to a combination of helical
turbulence (leading to the $\a$ effect), differential rotation (the $\Omega$
effect),  and turbulent diffusion to exponentiate an initial seed mean
magnetic field. The total magnetic field is split into a mean
component  and a fluctuating component, and the rate of growth of the 
mean field is sought. The
mean field grows on a length scale much larger than the outer scale 
of the turbulent velocity, with a growth time much larger than the 
eddy turnover time at the outer scale.  A combination of 
kinetic and current helicity provides a statistical correlation of 
small scale loops
favorable to exponential field growth.  Turbulent diffusion is needed to
redistribute the amplified mean field rapidly to ensure a net 
mean flux gain inside the system of interest.
Rapid growth of the fluctuating field necessarily accompanies the 
mean-field dynamo.  Its impact upon the growth of the mean
field, and the  impact of  the mean field itself on its own growth are
controversial.

The controversy results because  
Lorentz forces from the growing magnetic field react back on and
complicate the turbulent  motions driving the field growth  (e.g. Cowling
1959, Piddington 1981, Kulsrud \& Anderson 1992). It is tricky to
disentangle the back reaction of the mean field from that of the 
fluctuating field. Analytic studies and numerical
simulations seem to disagree as to the extent to which the dynamo
coefficients are suppressed by the back reaction of the mean field. 
Some numerical studies (e.g.\
\cite{CAT91}, \cite{VAI92},
\cite{CAT94}, \cite{CAT96})
and analytic studies (\cite{GRU94}, \cite{BHA95}, \cite{KLE95}) 
 argue that the suppression of  
$\a$ takes the form  
$\a\sim \a^{(0)}/(1+R_M^p{B}^2/v_0^2)$
where $\a^{(0)}$ is the value
of $\a$ in the absence of a mean field,  $R_M$ is the magnetic
Reynolds number, $\bbB$ is the  mean field in velocity units, $v_0$
is the {rms} turbulent velocity, and $p$ is a number of order 1.
Such a strong dependence on
$R_M$ would 
prevent astrophysical dynamos from working, as $R_M$ is usually $\gg 1$.  

Other numerical studies (Brandenburg \& Donner 1997) and
analytic studies (e.g. Kraichnan 1979; Field et al., 1999, Chou \& Fish 1999)
suggest  that $p=0$, so
$\a\sim \a^{(0)}/(1+\bB^2/v^2_0)$ in the fully dynamic regime.
In particular Field et al. (1999)
considered an expansion in 
the mean magnetic field (see also Vainshtein \& Kitchatinov 1983; 
Montgomery \& Chen 1984; Blackman \& Chou 1997), 
and were able to
derive the effect of the nonlinear back reaction on $\a$ in the
case for which $\nabla\bbB=0$.
Their result is expressed in terms  only of the
difference between the zeroth-order  kinetic and current helicities.  
They find that  $R_M$ does not enter strongly, except possibly by
suppressing the difference between the zeroth-order helicities, an effect
which cannot depend upon $\bbB$ and is not therein constrained.

Blackman \& Field (1999) have shown that some of the analytic approaches
(e.g. Bhattacharjee \& Yuan 1995;
Gruzinov \& Diamond 1994)
which employ a use of Ohm's law dotted with the fluctuating component
of the magnetic field, do not distinguish between turbulent quantities  
of the base (zeroth-order) state and quantities which are of higher
order in the mean field.  This distinction is important. When it is
made, many arguments for suppression via such approaches 
fall through.  Note that Blackman \& Field (1999) do not prove that the dynamo
survives  back reaction as a result of their considerations, 
only that some analytic approaches to the problem can be challenged.   

Despite this challenge, the apparent result of extreme
$\alpha$ suppression is  seen in the simulation of Cattaneo
\& Hughes (1996). These authors externally force the turbulence, 
imposing periodic boundary conditions and 
a uniform mean field, and find that the suppression of $\a$ 
involves $R_M$ in the form given above.  By contrast, 
the simulation of Brandenburg \& Donner (1997) suggests that
an $\a-\Omega$ dynamo may in fact be operative in an accretion disc
whose turbulence is self-generated by a shearing instability, without
$R_M$ entering the suppression.
The latter simulation does not employ periodic boundary conditions and
allows gradients in mean fields.

In this paper we show that the suppression of $\a$  depends 
crucially on the boundary conditions.
We find that the mean quantities are defined by averaging over a periodic 
box $\a$  
has an upper limit that depends on a factor of $R_M^{-p}$, with $p\sim
1$.  
In the presence of mean field gradients and non-periodic boundary
conditions, however, we find that the
upper limit on the dynamo coefficients is significantly larger,
and $R_M$ is not involved.  
The small upper limit in the periodic box case does not 
represent a dynamical suppression but rather
an apparent suppression which is just a results from the boundary 
conditions.  The results herein may be a step toward 
resolving controversies surrounding numerical suppression experiments.

Central to the discussion is the equation for the time evolution
of magnetic helicity.  This equation was also employed by
Seehafer (1994), who derived a suppression of $\a$  apparently
consistent with that of Keinigs (1983) (and qualitatively 
consistent with the Cattaneo \& Hughes (1996) simulation).  
The techniques of these additional 
two papers are different and one should note that   
they do not separate zeroth from higher order quantities.

Section 2 reviews the basic formalism of the dynano coefficient
expansion in orders of $\bbB$. 
Section 3 shows that constraints on the magnitude of the EMF (and thus $\a$)
results from dotting Ohm's law for the fluctuating electric field
with the fluctuating magnetic field, taking the average and expanding
to second order in the mean magnetic field.  The results depend
on the boundary conditions.
For a periodic box, the upper limit on $\a$ is too small
for a dynamo to work in practice, but this does not represent a dynamical
suppression.
In section 4 we interpret the results in terms of helicity flow 
and we discuss implications with respect to 
previous studies.  Section 5 is the conclusion.  

\medbreak
\centerline {\bf 2. Basic Formalism}

The basic formalism employed herein is discussed in 
Field et al. (1999) and Blackman \& Field (1999).
The formalism combines some aspects of the standard textbook treatment
(e.g. Moffatt 1978)
with the modification that fluctuating quantities
are divided into (1) 
a zeroth-order turbulent base state whose correlations are 
homogeneous, stationary and isotropic (though not necessarily 
mirror-symmetric!), 
and (2) a contribution which depends on the presence
of a non-zero mean field (see also Vainshtein \& Kitchatinov 1983; 
Kitchatinov et al. 1994). This higher-order contribution is 
definitely not isotropic and not necessarily homogeneous or stationary.

More specifically, 
the induction equation describing the magnetic field evolution is
\begin{equation}
\partial_{t}{\bf B} =\curl(\bfV\ts\bf B) + 
\la\nabla^{2}{\bf B},
\label{INDUCTION}
\end{equation}
where $\la=\eta c^2/4\pi$ is the magnetic viscosity, corresponding
to resistivity $\eta$.
Here $\bf B= \bfb +\bbB$ is the magnetic field 
  in velocity units, obtained by dividing by $\sqrt{4\pi\rho}$,
and $\bfb$ and $\bbB$ are the fluctuating and mean components of
${\bf B}$ respectively.  We assume incompressibility.

The equation for the mean field
derived by averaging  (\ref{INDUCTION})   is
\begin{equation}
\partial_{t}\bbB =
\curl\lb\bfv \ts \bfb\rb
-\bbV\cd\nabla\bbB
+\bbB\cdot\nabla\bbV
+\la\nabla^{2}\bbB.
\label{BBAR}
\end{equation}
The equation for $\bfb$ is given by subtracting (\ref{BBAR}) 
from (\ref{INDUCTION}), which gives
\beq 
\partial_t \bfb=\curl(\bfv\ts\bfb)-\curl\lb\bfv\ts\bfb\rb
+\curl(\bfv \ts \bbB)+\curl (\bbV\ts \bfb).
\label{fluct}
\ee

The term $\bbB\cdot \nabla\bbV$ in (\ref{BBAR})
describes the $\Omega$-effect  of differential rotation, and will not be
discussed further here, while the term $\bbV\cd\nb\bbB$ can be eliminated
by changing  the frame of reference to one moving with $\bbV$; both terms
will be ignored in what follows. 
The dynamo theorist must find the
dependence of the  turbulent EMF $\lb\bfv\ts\bfb\rb$ on $\bbB$ so that
(\ref{BBAR}) can be solved.

In the absence of the mean velocity fields, all mean vectors
can be written in terms of the mean magnetic field.
In particular, we have 
\begin{equation}
\lb\bfv\ts\bfb\rb=
\alpha_{ij} \bB_j-\beta_{ijk}\partial_j\bB_k + \ga_{ijkl}{\rm O}({\overline B}/R^2)+...,
\label{2EMF}
\end{equation}
where $\a_{ij}$, $\beta_{ijk}$ and $\ga_{ijk}$ are 
explicit functions of correlations of turbulent quantities, but
can depend implicitly on $\bbB$ (Moffatt 1978) through their
dependence on the induction equation for the fluctuating field.
The order at which there
is no implicit dependence on $\bbB$ is the zeroth-order base state
(see Field et al 1999).
The expansion order parameter is $|\bbB|/|\bfb^{(0)}|$, which is
indeed $<<1$ for the early dynamo evolution, and $<1$ in the 
Galaxy at present.  In particular, we have 
$\bfb=\bfb^{(0)}+\sum_n\bfb^{(n)}$, and similarly for $\bfv$, where
$\sum_n\bfb^{(n)}<\bfb^{(0)}$ and 
$n$ indicates the number of powers of $|\bbB| / |\bfb|$. 
The zeroth order base state correlations are composed of products 
of $\bfb^{(0)}$ and $\bfv^{(0)}$ and have no dependence on the mean field. 
The zeroth order base state is taken to be homogeneous and isotropic--the
violation of isotropy comes from the contributions due to higher
order fluctuating quantities, whose isotropy is broken by the mean field.
Note that the zeroth order state need not be reflection invariant, and
it is important for dynamo theory that it is not.

Correlations between higher order quantities can be reduced to
correlations of zeroth order quantities times the respective
products of $n$ linear functions of $\bbB$.  
Thus for example, $\bfb^{(2)}$ is the anisotropic component of the fluctuating
magntetic field which depends on two powers of $\bbB$, and is found
by twice iterating terms like $\bfb\cdot\bbB$ in the induction 
equation to obtain an approximate solution in terms of $\bfb^{(0)}$
and $\bfv^{(0)}$.

To zeroth order, the $\alpha$ tensor can be written 
\begin{equation}
\alpha_{ij}^{(0)} =\a^{(0)}\de_{ij},
\label{3EMF}
\end{equation}
which highlights the isotropy of this zeroth-order quantity.
In our previous work (Field et al.\ 1999) we have used the induction
equation for the fluctuating components of the magnetic field
and the Navier-Stokes equation for the fluctuating velocity  to
find the form of $\a$ in terms of correlations of the 
zeroth-order products (see also Blackman \& Chou 1997).  
Calculating the turbulent EMF in the absence of gradients of $\bbB$, 
to first order in $\bB$, 
gives $\lb\bfv\ts\bfb\rb^{(1)}=\a^{(0)}\bbB$, where
$\a^{(0)}$ is the sum of kinetic and current helicities associated
with the zeroth order state, namely
\begin{equation}
\begin{array}{r}
\a^{(0)} = 
- {1\over 3}\lrb{\lra{\bfv^{(0)}(t)\int\cnt
\bfv^{(0)}(t')dt'}-\lra{\bfb^{(0)(t)}\int\cnt \bfb^{(0)}(t')dt'}}\\
\simeq - {1\over 3}t_c \lrb{\lra{\bfv^{(0)}\cnt
\bfv^{(0)}}-\lra{\bfb^{(0)}\cnt \bfb^{(0)}}},
\end{array}
\label{2.4}
\end{equation}
where $\bfv$ is the turbulent velocity and $t_c$ defined as the 
correlation time of the scale of the turbulence
which dominates the averaged quantity.
If we adopt a Kolmogorov energy spectrum 
(i.e. $kb_k^2,kv_k^2 \propto k^2E_k \propto k^{1/3}$), then it might appear
that the dominant contributions to the terms of (\ref{2.4}) come from 
large $k$. 
However, Pouquet {\it et al.}\ (1976) showed that
if the forcing is at the outer wavenumber $k_0 = L^{-1}$, 
most of the energy and helicity 
is concentrated there, and the turbulence for $k>3k_0$
is locked up into Alfv\'en waves which do not 
contribute to correlations.
It is therefore likely reasonable to assume that any helicity in
the zeroth-order state is concentrated near 
$k_0$, in which case $t_c\sim (v_0k_0)^{-1}\sim L/v_0$.

The first term in (\ref{2.4}) was first derived by Steenbeck, Krause, \&
R\"adler (1966). The second, current helicity, term in
\refp{2.4} was first derived by Pouquet et al.\ (1976); neither paper made
the necessary distinction between zeroth and higher-order quantities.

In the next sections we will not re-derive the 
form of $\a^{(0)}$ in
terms of $\bfb^{(0)}$ and $\bfv^{(0)}$;  instead we will  derive an
independent upper limit on $\a^{(0)}$ from 
the use of Ohm's law, the definition of the electric field in terms
of the vector potential, and the equation for magnetic
helicity evolution.  We will invoke the assumption that 
the zeroth-order  
base state is isotropic and homogeneous, and we will assume that
that all anisotropies and inhomogeneities of higher-order correlations
are due to mean fields.

We will need the Reynolds relations
(\cite{RAD80}), i.e. that derivatives with respect to $\bf x$ or $t$
obey $\partial_{t,{\bf x}}\lb { X_i X_j} \rb =
\lb\partial_{t,{\bf x}}({X_i X_j})\rb$ and $\lb
\overline{X}_{i}x_{j}\rb = 0$
where $X_i={\overline X}_i+x_i$ are
components of vector functions of $\bf x$ and $t$.  
For statistical ensemble means, these hold when  
correlation times  are small compared to the  
times over which mean quantities vary.  For spatial means,
defined by $\lb X_{i}({\bf x},t)\rb = V^{-1} 
\int X_{i} ({\bf x}+{\bf s},t) {\rm d}{\bf s}$,
the relations hold when the average is over a large enough
$V$ that $L \ll V^{1 \over 3}\le R \le D$, where $D$ is the size of the system
$R$ is the scale of mean field variation and
$L$ is the outer scale of the turbulence.  Note that the
scale of averaging is less than the overall system size.

\medbreak

\centerline{\bf 3. Constraints on the tubulent EMF for periodic and 
non-periodic boundary conditions}

\ni {\bf a. Constraint equations}

Let the electric field $\bfE$, like $\bfB$, be divided into a mean component
$\bbE$ and a fluctuating component $\bfe$.  
Ohm's law for the mean field is thus  
\beqn
\bbE= \lra{-c^{-1}\bfV \times \bfB +\eta \bfJ} 
= -c^{-1}\lra{\bfv\times \bfb}+\eta \bbJ \lab{mon6}
\eeqn
for the case $\bbV=0$, where ${\overline {\bf J}}$ is the current
density and $\eta$ is the resistivity.
We also have 
\beqn
\lb{\bfE \cdot \bfB}\rb 
=\bbE\cdot \bbB+\lra{\bfe\cdot \bfb} 
= -c^{-1}\lra{\bfv\times \bfb}\cdot \bbB+\eta \bbJ\cdot \bbB+
\lra{\bfe\cdot \bfb} 
\lab{mon21}
\eeqn
where we have used (\ref{mon6}).

A second expression for $\lb{\bf E}\cdot {\bf B}\rb$ also follows from 
Ohm's law without first splitting into mean and fluctuating components,
that is 
\beqn
\lra{\bfE\cdot \bfB}& = &\lra{-c^{-1} (\bfV\times \bfB )\cdot \bfB + \eta
\bfJ\cdot \bfB} \nn
&=& \eta \lra{\bfJ\cdot \bfB} = \eta \bbJ\cdot \bbB+\eta
\lra{\bfj \cdot \bfb} =\eta\bbJ\cdot \bbB \nn
&&+c^{-1} \lambda \lra{\bfb\cdot\nabla\times \bfb}.
\lab{mon22}
\eeqn
By substituting 
\refp{mon22} into  \refp{mon21}, we obtain 
\beq
c^{-1} \lra{\bfv\times \bfb} \cdot \bbB = -c^{-1} \la \lra{\bfb\cdot
\nb\times \bfb } +\lb{\bf e}\cdot {\bf b}\rb
, \lab{mon23}
\ee
an equation which will now constrain $\lra{\bfv\times \bfb}$. 
However, we must expand (\ref{mon23}) to second
order in $\bbB$ (as defined in section 2) 
to constrain the turbulent EMF $\lb\bfv \ts \bfb \rb$.
This is because to zeroth order, the left hand side of (\ref{mon23})
vanishes directly.  To first order, the left side would be 
$\lb\bfv \ts \bfb \rb^{(0)}{\bbB}$, but
$\lb\bfv \ts \bfb \rb^{(0)}=0$, since vector averages of 
zeroth order quantities vanish.
To second order in $\bbB$ then,  
\refp{mon23} implies that
\beq
c^{-1} \lra{\bfv\times \bfb}^{(1)}\cd \bbB= - c^{-1}\la\lra{\bfb\cnt \bfb}^{(2)} +\lb{\bf e}\cdot{\bf b}\rb^{(2)}.
\label{fri11a}
\ee 
Because $R_M>>1$, significant
limits on  $\lra{\bfv\times \bfb}^{(1)}$ and thus on $\a^{(0)}$
come from the $\lb{\bf e}\cdot\bfb\rb^{(2)}$ term above.
The result of Seehafer (1994) and Keinigs (1983) amount to the
(\ref{fri11a}) with the last term equal zero, but without distinguishing
the order in mean fields (i.e. without the superscripts).
We now focus on this last, term keeping in mind that it is second
order in mean fields.

Since $\lb{\bf e}\cdot\bfb\rb^{(2)}$ 
is second order in $\bfB$, its most general 
form will be expressible as a sum of terms which each involve
products of two types of quantities:
1. correlations of scalar or pseudoscalar products of zeroth order quantities 
and 2. quadratic scalar or pseudoscalar functions of $\bbB$.
Now   
$\lb{\bf e}\cdot\bfb\rb$ can be written as a sum a time derivative
and spatial divergence. 
Consider ${\bf e}$ in terms of the vector 
and scalar potentials ${\bf a}$ and $\phi$:
\begin{equation}
\bfe=-\nabla \phi -(1/c)\partial_t {\bf a}.
\label{1}
\end{equation}
Dotting with $\bfb=\curl{\bf a}$ and averaging we have
\begin{equation}
\lb \bfe\cdot \bfb\rb=-\lb \nabla \phi\cdot\bfb\rb 
-(1/c)\lb{\bf b}\cdot \partial_t{\bf a}\rb.
\label{2}
\end{equation}
After straightforward algebraic manipulation, application of Reynolds rules and $\nabla\cdot \bfb=0$, this equation implies
\begin{equation}
\lb \bfe\cdot \bfb\rb=- (1/2)\nabla\cdot  \lb \phi\bfb\rb +(1/2)\nabla \cdot\lb{\bf a}\ts {\bf e}\rb -(1/2c)\partial_t \lb{\bf a}\cdot \bfb\rb\equiv
-\partial_0 {\overline h}^{0} +\partial_i {\overline h}^{i}=\partial_{\mu}
{\overline h}^{\mu},
\label{3}
\end{equation}
where we have defined a helicity density 4-vector for fluctuating
quantities 
\begin{equation} 
[h_{0},h_i]=[(1/2c){\bf a}\cdot \bfb, (1/2)\cdot  ( \phi\bfb) 
-(1/2)({\bf a}\ts {\bf e})],  
\label{3aa}
\end{equation}
and the overbar is used, as always, to mean the same thing
as the brackets.  

\ni {\bf b. Constraints for periodic boundary conditions}

We now investigate the implications of (\ref{3})
for simulations of type performed by
Cattaneo \& Hughes (1996), where the brackets 
brackets are interpreted as a spatial average
over a periodic box.  Under these conditions,
there are two important consequences.
First, note that 
the second two terms of (\ref{3}) vanish upon conversion to surface 
integrals and we have 
\begin{equation}
\lb \bfe\cdot \bfb\rb=-(1/2c)\partial_t \lb{\bf a}\cdot \bfb\rb,
\label{3p}
\end{equation}
which is gauge invariant.
The second consequence of the periodic box is that 
$\partial_t \bbB=0$ for incompressible flows.
This follows simply from (\ref{BBAR}):
the last three terms of (\ref{BBAR}) would 
vanish as they are all surface integrals.
Using Reynolds rules and vector identities, the 
second term can be written 
$[\curl\lb\bfv\ts\bfb\rb]_j=
\lb\partial_i(b_i v_j)\rb -\lb\partial_i(v_i b_j)\rb$
which also vanishes by surface integration.

The  two consequences just discussed can be used  to show that
(\ref{3p}) vanishes for a periodic box, and thus
the only contribution to the right of (\ref{fri11a}) will come
from the first term on the right. 
To second order in mean quantities, assuming $\bfb(t=0)=0$
and that all times are far enough from $t=0$ such that 
$\bfb(t)$ does not correlate with any finite $\bfa(0)$, we have 
\beq
\lb{\bf a}\cdot \bfb\rb^{(2)}=
\int \lb\partial_{t'}\bfa^{(2)}(t')\cdot\bfb^{(0)}(t)\rb dt'
+\int \lb\partial_{t'}\bfa^{(1)}(t')\cdot\int\partial_{t''}
\bfb^{(1)}(t'')\rb dt'dt''
+\int\lb \bfa^{(0)}(t)\cdot \partial_{t'}\bfb^{(2)}(t')\rb dt'.
\label{aexp}
\ee
To express (\ref{aexp}) explicitly in terms of mean fields,
we use the equations of motion for $\bfb$ and $\bfa$.
The use of $\partial_t\bfb$ from (\ref{fluct})
for the last two terms of (\ref{aexp}) 
leads directly to contributions depending on products of the mean fields
$\bbB$ or $\bbV$  and turbulent quantities $\bfb$ and $\bfv$.
Consider now the equation for $\bf a$ which comes from uncurling
the equation for $\bfb$, 
namely
\beq 
\partial_t {\bf a}=(\bfv\ts\bfb)-\lb\bfv\ts\bfb\rb
+(\bfv \ts \bbB)+(\bbV\ts \bfb) + \nabla\theta,
\label{flucta}
\ee
where $\theta$ is an arbitrary scalar field.
When (\ref{flucta}) is used in (\ref{aexp})
in the first and second terms on the right of (\ref{aexp}), 
the periodic box nullifies the contribution from
$\nabla\theta$.  All other contributions depend only
on products of $\bfv$, $\bfb$, $\bbB$ and $\overline {\bf V}$.
Thus when $\overline {\bf V}=0$, the only remaining mean field 
is $\bbB$.  Thus for a periodic box, $\lb\bfa\cdot\bfb\rb^{(2)}$ must be 
second order in $\bbB$.
Then, when plugged into (\ref{3p}) the time derivative will act
on some quadratic function of $\bbB$ multiplied by
correlations of zeroth order.  Since the zeroth order quantities are time
independent, isotropic, and homogeneous, the function of $\bbB$ must
be a scalar, denoted $F$,  and we have 
\beq
\partial_t\lb\bfa\cdot\bfb\rb^{(2)}= \partial_t ({F}(\bbB)^{(2)}{\overline Q}_1^{(0)})
={\overline Q}_1^{(0)}\partial_t ({ F}(\bbB)^{(2)}) =0,
\label{result}
\ee
where ${\overline Q}_1^{(0)}$ is a scalar
or pseudoscalar correlation of zeroth order quantities. 
The last equality of (\ref{result}) follows form stationarity of zeroth
order quantities, and our proof that $\bbB$ is time independent
over the time scales of interest for the periodic box.
We therefore conclude that $\partial_t\lb\bfa\cdot\bfb\rb=0$
in (\ref{3p}). 
This result relates to the the fact that 
for a periodic box, there is no periodic mean vector field 
$\overline {\bf A}$ whose curl is everywhere 
equal to $\bbB$. The divergence of $\bbB$ is still equal to zero, 
so Maxwell's equations are satisfied, but $\bbB$ is the only
non-trivial mean field. 



Since in the Cattaneo \& Hughes (1996) 
simulation $\bbB=$constant in both space and
time, $\lb\bfv\ts\bfb\rb^{(2)}=\alpha ^{(0)}\bbB^2$. 
Using this, and (\ref{result}), (\ref{3p}) and  (\ref{fri11a}),  
we obtain
\beq
|\alpha^{(0)}|= {c^{-1} \la |\lra{\bfb\cnt\bfb}^{(2)}| \over \bB^2}.
\lab{fri37}
\ee
Field et al. (1999) showed that conclusions about $\a^{(0)}$ are also 
conclusions about $\a$ to all orders, by relating the
fully non-linear $\a$ to $\a^{(0)}$ and showing that in the limit
of large $\bB$, $\a$ is not catastrophically affected when $\bB$ is large.
Thus $\a^{(0)}$ is an upper limit to $\alpha$, and so the result
(\ref{fri37}) shows that $\a$ will be small when the brackets
indicate an average over a periodic box.  The important point
is that this is not a dynamical suppression from the backreaction
but a constraint on the magnitude of  
$\alpha ^{(0)}$ which is imposed by the boundary conditions.
Notice that it is a constraint on the zeroth order quantity,
and so it cannot represent the effect of  backreaction.


\ni {\bf c. Constraints for non-periodic boundary conditions}

If the averaging brackets are not over a periodic box, or if the
scale of the averaging is $<<$ than the overall scale size of the system,
then the divergence terms in (\ref{3}) 
do not vanish.  In addition,
the $\nabla \phi$ term in (\ref{flucta}) will contribute to
(\ref{3p}).  In this case, each term on the right of (\ref{3}) 
is not gauge invariant.  Thus, the only constraint we can make
on the magnitude of the right side of (\ref{3}) is on the
sum of all the terms together.  Writing down all possible
second order terms up to one spatial 
derivative in $\bbB$, we have 
\beq
c\lb \bfe \cdot\bfb\rb^{(2)}={\overline Q}_2^{(0)}\bbB^2 + 
{\overline Q}_3^{(0)}\bbB\cdot\curl\bbB + {\rm O}({\overline B}^2L^2/R^2)+...,
\label{result2}
\ee
where ${\overline Q}_2^{(0)}$ and ${\overline Q}_3^{(0)}$
are correlations of zeroth order averages, $L$ is the outer turbulent
scale and $R$ is the mean field variation scale.
The quantity ${\overline Q}_2^{(0)}$ must have units of velocity,
and thus be of maximum order $v_0$ since it depends only on turbulent
quantities.  The quantity ${\overline Q}_3^{(0)}$ must have dimensions 
of viscosity, and must of maximum $\sim v_0L$ since it too depends only 
on turbulent quantities.  The combination of terms
in (\ref{result2}) is the same form of the combination of terms
entering on the left side of (\ref{mon23}) which would result from using
(\ref{2EMF}).  That is, 
since $\lra{\bfv\times \bfb} \cdot \bbB = \alpha^{(0)}\bbB^2-\beta^{(0)}
\bbB\cdot\curl\bbB$, we have ${\overline Q}_1^{(0)}=\a^{(0)}$ 
and ${\overline Q}_2^{(0)}=-\beta^{(0)}$.
Thus for simulations in which the mean values are not taken over
a periodic box, there is certainly no a priori restriction on $\alpha^{(0)}$.
Since now $\a\le\a^{(0)}$, any simulation
result indicating suppression on $\alpha$ 
under these relaxed boundary conditions would indeed
be a dynamical suppression.  So far there are no simulations
which invoke such boundary conditions that show catastrophic
suppression (c.f. Brandenburg \& Donner 1997).

\centerline {\bf 4. Discussion} 

Section 3 shows that periodic boundary conditions
impose an upper limit on $\a$ that does not represent a dynamical suppression.
Non-periodic boundary conditions or a finite scale separation between
system size and mean field gradient scale
allow for a much higher upper limit on $\alpha$, namely
the kinematic limit. 
The dynamical backreaction is testable only in simulations of the latter type.

{\bf 4.1 Relation to magnetic helicity}

Here we point out a connection to magnetic helicity.
Repeating Eqns. (\ref{1}), (\ref{2}) and (\ref{3})
for the total $\bf E$ and $\bf B$ gives 
\beq
\lb{\bf E}\cdot {\bf B}\rb = 
{\overline{\bf E}}\cdot \bbB +\lb\bfe\cdot\bfb\rb
= {1\over 2}
\partial_\mu{H}^{\mu}={1\over 2}
\partial_\mu{\widetilde h}^{\mu}+{1\over 2}\partial_\mu{\overline h}^{\mu}\simeq 0,
\label{ref1}
\ee
where $H^{\mu}$ is the total magnetic helicity 4-vector (Field 1986)
defined exactly as in (\ref{3aa}) but with all fluctuating quantities replaced
by their total values.
Similarly, ${\widetilde h}^{\mu}$ is the helicity 4-vector associated
with the mean fields.  The last similarity in (\ref{ref1})
follows because 
$R_M >> 1$ ($\la \sim 0$) in the astrophysical plasmas of interest.
Using $\lb\bfe\cdot\bfb\rb=\partial_\mu {\overline h^{\mu}}$,  
Eqn. (\ref{ref1}) then shows that any non-negligible 
$\lb\bfe\cdot\bfb\rb$ requires a finite but opposite
$\partial_\mu {\widetilde h^{\mu}}$.


In general, for non-vanishing turbulent EMF, the $\lb\bfe\cdot\bfb\rb$
must be non-zero, and thus the  
4-divergences in (\ref{ref1}) cannot vanish separately.
Interestingly, 
when (\ref{ref1}) is integrated over the total volume inside and outside
of the rotator, and interpreted in terms of a flow of relative
magnetic helicity (Berger \& Field 1984), 
it can be shown that a working dynamo implies an associated
magnetic energy flow through the magnetic rotator of interest
which likely leads to an active corona 
(Field \& Blackman 2000; Blackman \& Field 2000).

{\bf 4.2 Implications for Previous and Future Studies} 

The use of periodic boundary conditions in simulations appears
to be unsuitable for testing the suppression of $\a$ in a real dynamo 
unless the scale of mean field variations is much smaller than the
scale of the periodicity.  If periodic boundary conditions are used,
one must also be careful about causality issues.  The scale separation should 
at minimum be large enough such that the Alfv\'en crossing time across
the box is longer than correlation time of the fluctuating
quantities, and possibly even longer than the time scale for mean
field variation.
Thus, the box could be periodic, but the dynamics
of interest would occur in a non-periodic sub-region.
The brackets which we have used to indicate averages, would thus
represent averages over this sub-region, not the entire volume.
Alternatively, the box could be non-periodic.

The numerical simulations of Cattaneo \& Hughes (1996) 
do not allow for any mean field gradients and employ 
periodic boundary conditions. 
The strong  $\a$ reduction seen there is consistent with
our suggestion that the suppression may not be  dynamical, but may
instead be a result of the boundary conditions.  In contrast, 
the shearing box accretion disk simulations of Brandenburg \& Donner (1997)
do employ non-periodic boundary conditions and allow mean 
field gradients.  Interestingly, they do 
find that something like a mean-field dynamo is operating
therein.  The limited suppression that they find does not involve $R_M$.

\centerline {\bf 5. Conclusions}


We have suggested that the cause for apparent 
$\alpha$ suppression in numerical simulations which use periodic
boundary conditions may not result from dynamics, but from
rather from a choice of boundary conditions.
If the boundary conditions enforce all mean field gradients and 
spatial divergences to vanish, then the upper limit on $\a$ is
given by (\ref{fri37}). For non-periodic boundary conditions
or a box with significant scale separation between the mean field and box
size, the upper limit on the turbulent EMF is given by the
kinematic value.  
This would be a consistent interpretation of the large suppression 
reported by Cattaneo \& Hughes (1996).
In contrast, Brandenburg and Donner (1997) interpret
disk simulations which use non-periodic boundary conditions
and do not find such a strong suppression.
In summary, our results are 
consistent with seemingly contradictory simulations.


Working dynamos in real astrophysical bodies (even in the kinematic
approximation) require mean field gradients and scale separations
between the overall system scale, mean field averaging scale, and fluctuating
scale.  
In order to disentangle boundary effects from dynamical ones, 
future simulations of $\alpha$ suppression 
should include non-periodic boundary conditions 
or allow the mean field to change over scales smaller than the size of
the overall box.  This is a challenging task.


Acknowledgements: G.B.F. acknowledges partial support from NASA grant NAGW-931.
E.G.B. acknowledges support from NASA grant NAG5-7034.


\begin{thebibliography}{}




\bibitem[Beck et al. (1996)]{BEC96} Beck, R. et al. 1996,  Ann.
Rev.  Astron. Astrophys., {\bf 34}, 155

\bibitem[Berger \& Field (1984)]{BF84} Berger, M. A. \& Field, G. B., 
1984, J. Fluid Mech. {\bf147}, 133.

\bibitem [Bhattacharjee \& Yuan (1995)]{BHA95} 
Bhattacharjee, A. \& Yuan Y., 1995, ApJ, 449, 739.




\bibitem[Blackman \& Chou (1997)] 
{BLA97} Blackman, E.G. \& Chou, T., 1997, ApJ {\bf 489}
L95

\bibitem[Blackman \& Field (1999)] 
{BLA97b} Blackman, E.G. \& Field, G.B., 1999, ApJ, in press.

\bibitem[Blackman \& Field (2000)] 
{BLA97b} Blackman, E.G. \& Field, G.B., 2000, MNRAS, submitted.



\bibitem[Brandenburg (1997)]{BRA97} Brandenburg, A. \& Donner, K.J., 1997, 
MNRAS {\bf 288}, L29.

\bibitem[Cattaneo (1994)]{CAT94} Cattaneo, F. 1994, ApJ, {\bf
434}, 200

\bibitem[Cattaneo \& Hughes (1996)] {CAT96}
Cattaneo, F.  \& D.W. Hughes, 
1996, Phys. Rev. E., 54, 4532.

\bibitem[Cattaneo \& Vainshtein (1991)]{CAT91} Cattaneo, F. \&
Vainshtein, S.I., 1991, ApJ, {\bf 376}, L21.

 \bibitem[Chou \& Fish (1999)]{CF99} Chou, H.S. \& Fish, V., 1999, ApJ,
submitted.

\bibitem[Cowling (1957)]{COW57}Cowling, T.G., 1957, {\sl Magnetohydrodynamics},
(New York: Interscience).

 
\bibitem[Field 1986]{FIE86}Field, G. B., 1986, Magnetic Helicity in
Astrophysics, in {\it Magnetospheric Phenomena in Astrophysics}, R. I.
Epstein \& W. C. Feldman, eds., AIP Conference Proceedings 144 (Los
Alamos: Los Alamos Scientific Laboratory, 1986), 324.

\bibitem[Field 1995]{FIE95}Field, G. B., 1995, in {\sl Proceedings of the 
International Conference on Plasma Physics}, ed.P. H. Sakanaka \& E. del Bosco
(Woodbury, New York: AIP Press), 416.

\ni Field G.B. \& Blackman E.G. 1999, in {\it Highly Energetic Physical Processes and Mechanism for Emisson from Astrophysical Plasmas}, IAU Symp.
195 eds. P.C.H. Martens and S. Tsuruta.

\bibitem[Field, Blackman, Chou (1998)]{FIE98} Field, G. B. \&
Blackman, E. G. \& Chou, H.,  1999, ApJ, 513 638.



\bibitem[Gruzinov \& Diamond (1994)]{GRU94}
Gruzinov, A. \&  Diamond P.H., 1994, PRL, 72, 1651.



\bibitem[Kitchatinov et al. (1994b)]{KIT94c} Kitchatinov, L. L.,
Pipin V.V. and R{\"u}diger, G., \& Kuker, M. 1994, Astron. Nachr., 
{\bf 315}, 157


\bibitem[Keinigs (1983)]{KEI83} Keinigs, R.K., 1983, 
Phys. Flu. {\bf 26} 2558.

\bibitem[Kleeorin et al., (1995)]{KLE95} Kleeorin, N., Rogachevskii, \& Ruzmaikin, A., 1995, 
A\&A {\bf 297}
159

\bibitem[Kleeorin et al., (1982)]{KLE82}
Kleeorin, N. \& Ruzmaikin, A., 1982, Magnetohydrodynamics, N2, 17.

\bibitem[Kraichnan (1979a)]{KRAICHNAN} Kraichnan, R. H. 1979, 
Phys. Rev., {\bf 113}, 1181

\bibitem[Kraichnan (1979b)]{KRA79b} Kraichnan, R.H, 1979, PRL, {\bf 42}, 1667

\bibitem[Krause \& R{\"a}dler (1980)]{KRA80} Krause, F., \&
R{\"a}dler, K.-H. 1980, {\sl Mean-Field Magnetohydrodynamics and
Dynamo Theory}, Oxford: Pergamon Press

\bibitem[Kulsrud \& Anderson (1992)]{KUL92} Kulsrud, R. M., \&
Anderson, S. W. 1992, ApJ, {\bf 396}, 606

\bibitem[Moffatt (1978)]{MOF78} Moffatt, H. K. 1978, {\sl Magnetic
Field Generation in Electrically Conducting Fluids}, Cambridge:
Cambridge University Press

\bibitem[Montgomery \& Chen (1978)]{MON78} Montgomery, D. \& Chen, H., 1984,
Plasma Physics \& Controlled Fusion, 26, 1189.

\bibitem[Parker (1979)]{PAR79} Parker, E. N. 1979, {\sl Cosmical
Magnetic Fields}, Oxford: Clarendon Press


\bibitem[Piddington (1981)]{PID81} Piddington, J. H. 1981, {\it
Cosmical Electrodynamics}, Malbar: Krieger

\bibitem[Pouquet et al. (1976)]{POU76} Pouquet, A., Frisch, U., 
and Leorat, J. 1976, JFM {\bf 77}, 321

\bibitem[R{\"a}dler (1980)]{RAD80} R{\"a}dler, K.-H. 1980, Astron.
Nachr., {\bf 301}, 101

\bibitem[Ruzmaikin et al. (1988)]{RUZ88} Ruzmaikin, A. A., 
Shukurov, A. M., and Sokoloff, D. D. 1988, {\sl Magnetic Fields of
Galaxies}, Dordrecht: Kluver Press

\bibitem[Seehafer (1994)]{SHE94} Seehafer N., 1994, Europhys. Lett.,
{\bf 27} 353.

\bibitem[Steenbeck et al. (1966)]{Ste66} Steenbeck, M., Krause, F., \&
R\"adler, K.H., 1966, Z. Naturforsch., 21a, 369.


\bibitem[Vainshtein 1998]{V98} Vainshtein S., 1998, PRL 80, 4879.

\bibitem[Vainshtein \& Cattaneo (1992)]{VAI92} Vainshtein, S.I. \& 
Cattaneo F., 1992,  ApJ, {\bf 393}, 165.

\bibitem[Vainshtein \& Kitchatinov (1983)]{VAI83} Vainshtein, 
S. I., and Kitchatinov, L. L. 1983, Geophys. Ap.
Fluid. Dyn., {\bf 24}, 273


\bibitem[Zeldovich et al. (1983)]{ZEL83} Zeldovich, YA. B., Ruzmaikin, A. A., 
and Sokoloff, D. D. 1983, {\sl Magnetic Fields in Astrophysics
}, New York: Gordon and Breach

\end{thebibliography}
\end{document}